\documentclass[english,manuscript]{aastex62}
\usepackage[T1]{fontenc}
\usepackage[latin9]{inputenc}
\usepackage{amsbsy}
\usepackage{amstext}
\usepackage{amssymb}
\usepackage{graphicx}
\usepackage{array}
\usepackage{bm}
\makeatletter
\usepackage{amsmath}

\makeatother

\usepackage{babel}
\begin{document}

\title{Validating time-distance helioseismic inversions for non-separable subsurface profiles of an average supergranule}

\author{Vedant Dhruv}

\affil{Department of Astronomy and Astrophysics, Tata Institute of Fundamental
Research, Mumbai - 400005, India}

\author{Jishnu Bhattacharya}

\affil{Center for Space Science, New York University Abu Dhabi, Abu Dhabi
- 129188, United Arab Emirates}

\author{Shravan M. Hanasoge}
\affil{Department of Astronomy and Astrophysics, Tata Institute of Fundamental
Research, Mumbai - 400005, India}

\affil{Center for Space Science, New York University Abu Dhabi, Abu Dhabi
- 129188, United Arab Emirates}

\begin{abstract}
    Supergranules are divergent 30-Mm sized cellular flows observed everywhere at the solar photosphere. Their place in the hierarchy of convective structures and their origin remain poorly understood  \citep{Rincon2018}. Estimating supergranular depth is of particular interest since this may help point to the underlying physics. However, their subsurface velocity profiles have proven difficult to ascertain. \citet{Birch2006SPD....37.0505B} had suggested that helioseismic inferences would benefit from an ensemble average over multiple realizations of supergranules due to the reduction in realization noise. \citet{2017A&A...607A.129B} used synthetic forward-modelled seismic wave travel times and demonstrated the potential of helioseismic inversions at recovering the flow profile of an average supergranule that is separable in the horizontal and vertical directions, although the premise of this calculation has since been challenged by \citet{Ferret2019}. In this work we avoid this assumption and carry out a  validation test of helioseismic travel-time inversions starting from plausible synthetic non-separable profiles of an average supergranule. We compute seismic wave travel times and sensitivity kernels by simulating wave propagation through this background. We find that, while the ability to recover the exact profile degrades based on the number of parameters involved, we are nevertheless able to recover the peak depth of our models in a few iterations where the measurements are presumably above the noise cutoff. This represents an important step towards unraveling the physics behind supergranules, as we start appreciating the parameters that we may reliably infer from a time-distance helioseismic inversion.
\end{abstract}

\section{Introduction}

Convection near the surface of the Sun is manifested at various spatial and temporal scales, with the dominant contribution in Dopplergram measurements coming from granules that are about 1.5 Mm in size and last for minutes \citep{Spruit1990ARA&A..28..263S,DelMoro2004A&A...428.1007D}. These  are thermally driven and correspond to the convective injection scale at the solar photosphere. Masked beneath the transonic flow velocities of the granules are several larger-scale features that have lower velocity amplitudes, but significantly exceed granules in spatial dimensions and longevity. Supergranules are one such pattern that are observed everywhere on the solar disk, with typical sizes of around $30\,\text{Mm}$ and lifetimes of around $1.5$ days (see review by \citet{Rincon2018} and references therein). The velocity corresponding to supergranules is predominantly horizontal, in contrast to the dominant vertical flows associated with granules. While granules are understood to be convective cells, a comprehensive understanding of the physical mechanism that results in supergranules and the preference for the specific scale has remained elusive thus far. The velocity profiles of supergranules measured at the surface --- with upflows at the cell center feeding divergent outflows that get converted into downflows at the edge --- hints at a convective origin, however this has been debated, with \citet{Rast2003ApJ...597.1200R} and \citet{Crouch2007ApJ...662..715C} suggesting that the supergranular pattern emerges from collective dynamic interactions smaller scales. A pioneering hypothesis by \citet{Simon1964ApJ...140.1120S} that supergranules resulted from recombination of He II is not supported by supegranule-scale numerical simulations of near-surface solar convection \citep{Ustyugov2010PhST..142a4031U,Lord2014ApJ...793...24L}. \citet{Featherstone2016ApJ...830L..15F} have speculated that the peak in horizontal velocity spectra corresponding to supergranules is a consequence of suppression of power at even larger scales by solar rotation.  \citet{Gizon2003} and \citet{Schou2003ApJ...596L.259S} had discovered an oscillatory nature to supergranules that they interpreted as manifestations of traveling-wave convection, although this has met with disagreement \citep{Rast2004ApJ...608.1156R, Gizon2004IAUS..223...41G, Hathaway2006ApJ...644..598H}.

One of the reasons that the physics behind supergranules has proven hard to pinpoint is that their subsurface profiles have been difficult to infer. Several prior attempts in this regard using techniques such as time-distance heliosemology \citep{Duvall1998, Zhao2003,Jackiewicz2008}, helioseismic holography \citep{Braun2004,Braun2007} and correlations in Fourier space \citep{Woodard2007} show significant differences based on the specific techniques used in the study. The depths up to which subsurface flows can be reliably inferred is severely constricted by noise and therefore a statistical study is required. Individual supergranules might be thought of as realizations arising from an underlying stochastic driving mechanism and while they might differ in their specifics, a reliable estimate about their mean profile might be inferred by analyzing data corresponding to an ensemble average. \citet{Duvall2010} averaged over Doppler measurements of thousands of supergranular cells to produce one such profile using data from the Helioseismic and Magnetic Imager \citep[HMI,][]{Schou2012}. Such an averaging achieves two improvements: firstly it cuts down statistical noise on helioseismic measurements such as wave travel times, and secondly, it averages over realizations of supergranules to yield a mean surface profile. \citet{Duvall2012} used averaged  center-annulus travel time differences and carried out helioseismic ray-theoretic forward modelling to come up with a plausible subsurface flow profile that was consistent with the surface velocity profile. Their analysis suggested a best-fit flow model that peaked at $2.3\,\text{Mm}$ below the solar surface and rapidly decayed below a depth of $4\,\text{Mm}$. This result was in contrast with the study by \citet{Hathaway2012ApJ...749L..13H}, who suggested that supergranules are expected to extend to depths comparable to their diameters, and might interact significantly with the near-surface shear layer and other deeper flows in the solar convective zone \citep{Hathaway2012ApJ...760...84H}. A more thorough analysis was later carried out by \citet{Degrave2015} who concluded that the shallow model presented by \citet{Duvall2012} was not consistent with forward modelling in the Born approximation, throwing the question open once again. \citet{Degrave2015} also raised doubts over whether such an averaging procedure produces self-consistent results.

The lack of a clear answer compels one to go back to the basis of the inversion algorithm and validate it using synthetic flow profiles. Authors such as {\citet{Zhao2003,Zhao2007ApJ...659..848Z} have explored the accuracy of helioseismic inference using ray-theoretic travel time measurements, while \citet{Svanda2013ApJ...775....7S,Dombroski2013,Hanasoge2014,Bhattacharya2016} have carried out similar exercises in the Born approximation. Such an inversion typically proceeds by constructing forward-modelled travel times using a plausible model of subsurface flow in the Sun, comparing the surface measurements with those obtained from a solar model bereft of flows, using the differences in measurement to infer the flow velocities in the solar interior and finally comparing the flow profile thus obtained with the actual model that was used in the first stage. \citet{Zhao2007ApJ...659..848Z} had used the regularized least-square (RLS) technique and demonstrated that independent inversion for components of the flow velocity are often unable to decouple the contribution towards the travel time arising from the horizontal and vertical components (cross-talk), and assumptions such as mass-conservation that impose interrelations between them might alleviate this issue. The technique of subtractive optimally localised averaging \citep[SOLA,][]{Pijpers1992A&A...262L..33P,Jackiewicz2007,Jackiewicz2008} allows one to selectively independently invert for components of flow velocity. \citet{Svanda2011} validated a SOLA inversion for subsurface flows by suppressing cross-talk between components, and were able to successfully recover the velocity profile to a depth of $3.5\,\text{Mm}$ below the photosphere, and suggested that further improvements in signal-to-noise might be possible through ensemble averaging. Subsequently, attempts to validate helioseismic inversions using a synthetic mass-conserving average supergranule profile were carried out by \citet{Dombroski2013} using RLS inversion for travel times including realization noise, and by \citet{Hanasoge2014,Bhattacharya2016} without including realization noise. The result obtained by \citet{Dombroski2013} was broadly consistent with those by \citet{Zhao2007ApJ...659..848Z} and \citet{Svanda2013ApJ...775....7S}, in that inferential ability in depth was limited by noise. Further, \citet{Svanda2015} showed that the exact profile obtained depends on how strongly the solution is regularized. Inversion in the absence of noise is expected to be able to recover flow profiles to a greater depth than permitted otherwise. However, the results obtained by \citet{Hanasoge2014} and \citet{Bhattacharya2016} indicated that their scheme was unable to converge to the correct model, attributed by the authors to the high dimensionality of the parameter space. A modified approach was attempted by \citet{2017A&A...607A.129B}, who} were able to recover the depth profiles of supergranules by assuming a flow derived from an azimuthally symmetric stream function separable in radial and angular coordinates. The soundness of this assumption has since been contested by \citet{Ferret2019}, who demonstrated that surface measurements of flow velocities of a mass-conserving average supergranule are not consistent with separable models that have been used by various authors. In light of this, we extend the analysis by \citet{2017A&A...607A.129B} to a wider range of synthetic non-separable supergranule models. This leads to a significant increase in the number of parameters required to describe the model --- from tens to thousands --- and therefore, the present work is much more general than the previous analysis by \citet{2017A&A...607A.129B}. We show that while the technique does not necessarily reproduce the exact profiles, we are still able to estimate the depth to which models extend, and that this depth of the supergranule is recovered in the first few iterations where the travel time misfit is presumably above the noise cutoff. This is an encouraging result that can help us narrow down on the parameters that we may reliably infer from such an inversion procedure. 

\section{Supergranule Model}

\subsection{Kinematic description}

We superimpose a temporally stationary $2$D supergranular flow model on a convectively stabilized version of Model S \citep{ChristensenDalsgaard1996} as devised by \citet{Hanasoge2006}. The spatial extent of supergranules in the horizontal direction --- as observed on the solar photosphere --- is much smaller in comparison to the solar radius $R_{\odot}$, allowing us to ignore the curvature of the Sun and carry out our analysis in Cartesian coordinates $(x,y,z)$. We assume that the average supergranule is azimuthally symmetric, so we may limit our study to the $x-z$ plane. This assumption about isotropy does not strictly hold true \citep{Langfellner2015}, but is good enough for the present purpose. We assume that the axis of the supergranule coincides with the $z$-axis of our coordinate system. In the limit of the colatitude $\theta\rightarrow0$, Cartesian coordinates are related to spherical ones through $x\approx R_\odot\theta\cos\phi$ and $y\approx R_\odot\theta\sin\phi$. The imposition of $y=0$ necessitates $\phi=0$ or $\phi=\pi$ and so we may readily identify $x>0$ with $\phi=0$ and $x<0$ with $\phi=\pi$. We also impose a periodicity in $x$ over the length scale $L_x$ of our computational domain, so the coordinate $x$ takes values in $[-L_x/2,L_x/2)$ with $x=0$ coinciding with the center of the supergranule. The vertical coordinate $z$ denotes height above the solar surface and has zero value at the surface, negative values in the solar interior and positive values above the surface. Physical quantities of the solar model such as density profile $\rho(z)$, sound speed $c(z)$ and gravitational acceleration $\mathbf{g}=-g(z)\mathbf{\hat{z}}$ are stratified along $z$.  In further analysis we shall suppress the explicit coordinate dependence of the physical parameters wherever it is unambiguous.

We require the flow velocity $\mathbf{v}\left(\mathbf{x}\right)$ corresponding to the supergranule to satisfy the continuity equation

\begin{equation}\label{continuity_eqn}
\bm{\nabla}\cdot(\rho\mathbf{v})=0.
\end{equation}

This implies that the velocity field may be derived from an associated stream function $\bm{\psi}\left(\mathbf{x}\right)=\psi(x,z)\hat{y}$, through
\begin{equation}\label{velocity_from_psi}
\mathbf{v}=\frac{1}{\rho}\bm{\nabla}\times(\rho c\,\bm{\psi}).
\end{equation}

We note that this choice of a stream function directed along $\hat{y}$ might seem counter-intuitive given the azimuthal symmetry of the supergranule model. This is resolved by noting that we are primarily interested in recovering the flow profiles, and the specific choice of a stream function would not matter as long as it leads to similar velocity fields.

\begin{figure*}[t]
\centering
\includegraphics[width=18cm,height=4.7cm]{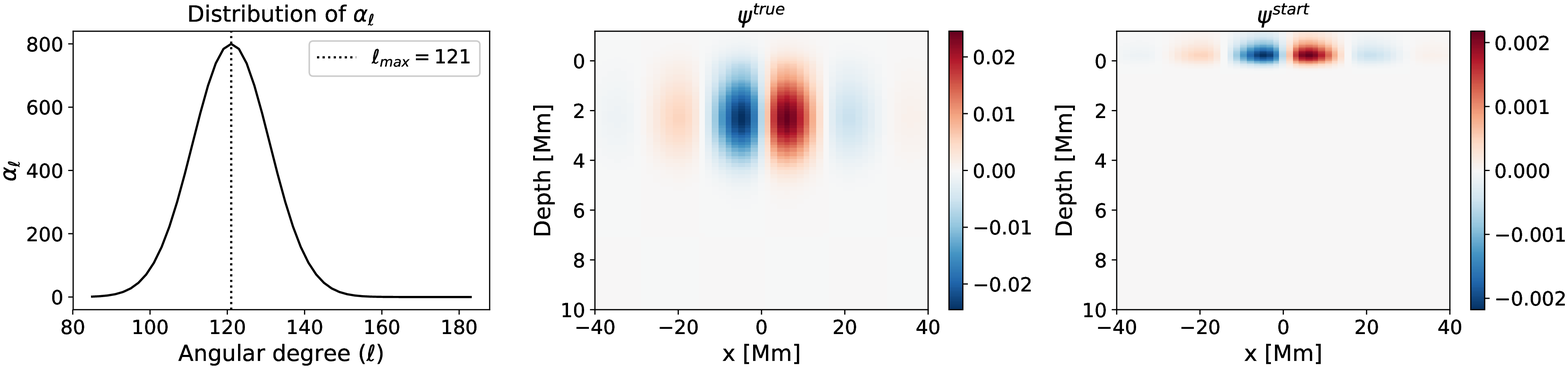}
\caption{\textit{Left panel}: Gaussian distribution that indicates the contribution of each angular degree to the flow model, \textit{middle and right panel}: True and starting stream function for SG$(\ell_4)$ respectively.}
\label{ell_distribution_and_true_and_starting_model}
\end{figure*}

We choose a non-separable (in $x$ and $z$) supergranule model by considering the stream function to be a weighted summation of separable profiles peaking at different depths and having distinct horizontal scales. The individual components have a form similar to that proposed by \citet{Duvall2012}, except the Bessel function for the horizontal variation in their analysis is replaced by normalized Legendre polynomials, defined as $\tilde{P}_\ell\left(x\right) =\sqrt{\left(2\ell+1\right)/2}\,P_\ell\left(x\right)$ where $P_\ell\left(x\right)$ represents the standard Legendre polynomial of degree $\ell$ for an argument $x$. The model for the supergranular stream function is
\begin{equation}\label{psi}
\psi\left(x,z\right)=\sum_{\ell}\alpha_{\ell}\frac{v_{0}}{ck}\tilde{P_{\ell}}\left(\frac{x}{L_x/2}\right)\exp\left(-\frac{(z-z_{\ell})^{2}}{2\sigma_{\ell}^{2}} - \frac{|x|}{R}\right),
\end{equation}
where $\alpha_{\ell}$ determines the contribution of each term towards the total stream function. We choose a Gaussian distribution peaking at $\ell=121$
\citep[corresponding to the size of a globally averaged supergranule as estimated from its power spectrum by][]{Williams2014} for the coefficients. The left panel of Fig. \ref{ell_distribution_and_true_and_starting_model} depicts this distribution for a particular model (SG($\ell_{4}$)). We consider odd $\ell$ only, which ensures the stream function goes to zero at $x=0$ and choose the constant parameters to be $R=10$ Mm and $k=2\pi/30$ Mm$^{-1}$. Henceforth, we will denote this flow model as the true model and label its parameters with the superscript ``true''. The iteratively updated flow model will be labelled ``iter''.

In our analysis, we consider two cases: Case $1$ comprises five supergranular models with increasing number of horizontal scales in the model definition. We wish to observe the efficacy of the inversion scheme as the number of parameters to be inverted for are progressively increased. In Case $2$, we examine four models peaking at different depths. Model parameters are listed in Table \ref{model_parameters} and are chosen so as to obtain a single-celled supergranule which produces surface velocities compatible with observations. The Gaussian distribution of $\alpha_{\ell}$ is characterized by the peak $\ell$ value, the standard distribution $\sigma_{\ell}$, amplitudes and cut-off $\ell$ values, $\ell_{min}$ and $\ell_{max}$. We list surface and peak velocities and supergranule depth (Eq. \eqref{supergranule_depth_def}) of the models in Table \ref{velocities and depth}.

\begin{table*}[t]
\centering
    \begin{tabular}{|| c c c >{\centering}p{2.5cm} c >{\centering}p{1.5cm}  >{\centering}p{3.4cm} >{\centering}p{4cm}||}
    \hline
         Model & $\ell_{min}$ & $\ell_{max}$ & Number of $\ell$s in true model & $\sigma_{\ell}$ & $v_{0}$ & Range of depths for angular degrees (Mm) & Total number of parameters for inversion \tabularnewline
         \hline
         SG$\ell_1$ & 119 & 123 & 3 & 1 & 350 & $2.5$ to $5$ & 69 \tabularnewline \hline
         SG$\ell_2$ & 115 & 127 & 7 & 1 & 350 & $2.5$ to $5$ & 161 \tabularnewline \hline
         SG$\ell_3$ & 109 & 137 & 15 & 4 & 350 & $2.5$ to $5$ & 285 \tabularnewline \hline
         SG$\ell_4$ & 85 & 183 & 50 & 10 & 550 & $2.6$ to $5$ & 950 \tabularnewline \hline
         SG$\ell_5$ & 41 & 239 & 100 & 25 & 300 & $2.5$ to $5$ & 1900 \tabularnewline \hline
         SG$d_1$ & 69 & 227 & 80 & 30 & 200 & $2$ to $4$ & 2400 \tabularnewline \hline
         SG$d_2$ & 69 & 227 & 80 & 30 & 250 & $3.4$ to $6$ & 2280 \tabularnewline \hline
         SG$d_3$ & 69 & 227 & 80 & 30 & 300 & $5.2$ to $8.2$ & 2400 \tabularnewline \hline
         SG$d_4$ & 69 & 227 & 80 & 30 & 350 & $6.6$ to $10$ & 2400 \tabularnewline \hline
    \end{tabular}
    \caption{Model parameters for models SG($\ell_1$) - SG($\ell_5$) and SG($d_1$) - SG($d_4$)}
    \label{model_parameters}
\end{table*}

\begin{table*}[t]
    \centering
    \begin{tabular}{||c c >{\centering}p{2cm} c >{\centering}p{2cm} c||}
    \hline
         Model & Max $v_{x}$ [m/s] & Max $v_{x}$ at surface [m/s] & Max $v_{z}$ [m/s] & Max $v_{z}$ at surface [m/s] & Peak depth (Mm) \tabularnewline
         \hline
         SG$\ell_1$ & 634 & 271 & 276 & 8 & 3.3 \tabularnewline
         \hline
         SG$\ell_2$ & 590 & 216 & 256 & 6 & 3.4 \tabularnewline
         \hline
         SG$\ell_3$ & 432 & 264 & 116 & 7 & 2.1 \tabularnewline
         \hline
         SG$\ell_4$ & 588 & 402 & 153 & 10 & 2.3 \tabularnewline
         \hline
         SG$\ell_5$ & 600 & 367 & 117 & 7 & 2.2 \tabularnewline
         \hline
         SG$d_1$ & 547 & 363 & 117 & 8 & 2.1 \tabularnewline
         \hline
         SG$d_2$ & 430 & 358 & 146 & 9 & 3.3 \tabularnewline
         \hline
         SG$d_3$ & 355 & 253 & 175 & 6 & 5.2 \tabularnewline
         \hline
         SG$d_4$ & 328 & 247 & 205 & 6 & 6.6 \tabularnewline
         \hline
    \end{tabular}
    \caption{Surface velocities, peak velocities and supergranule depth (Eq. \eqref{supergranule_depth_def}) for models SG($\ell_1$) - SG($\ell_5$) and SG($d_1$) - SG($d_4$)}
    \label{velocities and depth}
\end{table*}

\subsection{Basis decomposition of supergranular flow model}

The success of a nonlinear iterative inversion scheme depends on the
number of parameters that are being inverted for. This is reflected
in the results of \citet{Hanasoge2014} and \citet{Bhattacharya2016},
where a full-waveform inversion of the supergranule stream function
at every spatial point of the grid failed to converge to the true
model ($\sim10^{5}$ parameters for inversion). However, \citet{2017A&A...607A.129B} were successful in recovering the vertical profile of a $2$D separable supergranular flow model by decomposing the vertical dependence of the stream function on a spline basis. This approach enabled them to greatly reduce the number of parameters for inversion. We adopt a similar strategy and project the vertical profile of each term of
the stream function (Eq. \eqref{psi}) onto B-splines,
\begin{equation}
\begin{split}\psi^{\text{true}}(\mathbf{x}) & =\sum_{\ell}\alpha_{\ell}g_{\ell}(z)f_{\ell}(x)\\
 & =\sum_{\ell}\sum_{i=0}^{N-1}\alpha_{\ell}\beta_{i}^{\text{true}}B_{i}(z)f_{\ell}(x)\\
 & =\sum_{\ell}\sum_{i=0}^{N-1}c_{i\ell}^{\text{true}}B_{i}(z)f_{\ell}(x),
\end{split}
\label{psi_true_basis_decomposition}
\end{equation}
where $\beta_{i}^{\text{true}}$ represent the B-spline coefficients corresponding to the cubic B-spline $B_{i}(z)$. The B-spline functions are ordered with increasing $z$ value, i.e., $i=0$ corresponds to the B-spline function that peaks near the lower cutoff (this is chosen so that we can represent the flow model reliably and it is sufficiently below the turning point of the $p_{4}$ ridge) and $i=N-1$ corresponds to the B-spline function peaking closest to the upper-most vertical
coordinate in the grid. Similarly, the iterative model can be written as,
\begin{equation}\label{psi_iter_basis_decomposition}
\begin{split}\psi^{\text{iter}}(\mathbf{x}) & =\sum_{\ell}\sum_{i=0}^{N-1}\alpha_{\ell}\beta_{i}^{\text{iter}}B_{i}(z)f_{\ell}(x)\\
 & =\sum_{\ell}\sum_{i=0}^{N-1}c_{i\ell}^{\text{iter}}B_{i}(z)f_{\ell}(x).
\end{split}
\end{equation}
We consider the starting model of our inversion to have zero velocity below the surface and the same values as the true model at and above the surface, which are chosen to be commensurate with photospheric supergranule velocity measurements \citet{Duvall2010}. We achieve this by splitting the set of B-spline coefficients for every term in the stream function into two groups corresponding to those above and below the surface. Since the same set of knots are utilized for each term in the stream function expansion, the corresponding B-splines are the same as well. Consequently, the B-spline coefficient of index $m$ that has the maximum contribution near the surface is identical for all terms,
\begin{equation}\label{psi_above_and_below surface_basis_decomposition}
\begin{split}\psi^{\text{iter}}(\mathbf{x}) & =\sum_{\ell}\sum_{i=0}^{m-1}c_{i\ell}B_{i}(z)f_{\ell}(x)+\sum_{\ell}\sum_{i=m}^{N-1}c_{i\ell}B_{i}(z)f_{\ell}(x)\\
 & =\sum_{\ell}\sum_{i=0}^{m-1}c_{i\ell}B_{i}(z)f_{\ell}(x)+\sum_{\ell}\sum_{i=m}^{N-1}c_{i\ell}^{\text{surf}}B_{i}(z)f_{\ell}(x),
\end{split}
\end{equation}
where $c_{i\ell}^{\text{surf}}=\alpha_{\ell}\beta_{i}^{\text{surf}}$ are the coefficients of the true model peaking at and above the surface. The starting supergranule model is written as,
\begin{equation}\label{psi_start}
\psi^{\text{start}}(\mathbf{x})=\sum_{\ell}\sum_{i=m}^{N-1}c_{i\ell}^{\text{surf}}B_{i}(z)f_{\ell}(x).
\end{equation}
The true and starting flow models for the case SG($\ell_{4}$) are shown in the middle and right panels of Fig. \ref{ell_distribution_and_true_and_starting_model} respectively.

\section{Inversion Methodology}
We apply a full-waveform inversion technique to solve the inverse problem along the lines of \citet{Hanasoge2014}. A full-waveform inversion typically proceeds by optimizing model parameters to fit the entire measured wave field. We choose a simpler variant of this, and instead of fitting the time-dependent wave amplitude we compute the wave travel time following \citet{Gizon2002} at various points just above the solar surface, and try to fit these by iteratively updating our model of the supergranule. We use the publicly available code SPARC \citep{Hanasoge2007} to simulate wave propagation through supergranules on a Cartesian grid that spans $800$ Mm horizontally over $512$ pixels and extends from $1.18$ Mm above the surface to $137$ Mm beneath it over $300$ pixels spaced equally in acoustic distance. We choose eight sources at $150$ km below the photosphere at different horizontal locations that fire independently of each other, and the wave propagation from each source is tracked in separate simulations. Since the sources fire separately, we have eight different simulations that run in parallel for $4$ solar hours. 

Seismic waves in the Sun may be described in terms of their spatio-temporally varying displacement amplitude $\bm{\xi}(\mathbf{x},t)$ that evolves according to
\begin{equation}\label{wave_equation}
\rho\partial_{t}^{2}\bm{\xi}+2\rho\mathbf{v}\cdot\bm{\nabla}\partial_{t}^{2}\bm{\xi}=\bm{\nabla}(c^{2}\rho\bm{\nabla}\cdot\bm{\xi}+\bm{\xi}\cdot\bm{\nabla}p)+\mathbf{g}\bm{\nabla}\cdot(\rho\bm{\xi})+\mathbf{S},
\end{equation}
where $\mathbf{S}(\mathbf{x},t)$ represents sources that are producing waves and $\mathbf{v}$ represents the flow field associated with a supergranule. The vertical wave velocity is subsequently measured at a set of \textit{receivers} located at a height of $200$ km above the surface, and spread over a wide range of horizontal coordinates. We compute travel times between all source-receiver pairs for both true and iteratively updated supergranule models. We employ ridge-filters to isolate $f-p_4$ and measure travel times for each individual radial order and track them separately. 
We refer to the source-receiver travel time for a specific ridge as $\tau_{s,r,\text{ridge}}$, and we shall label the travel times with an appropriate superscript to indicate whether they are measured in the simulation with the true model or the one that we update. We combine these travel times into one misfit function $\chi$, defined as 
\begin{equation}
\chi=\frac{1}{2}\sum_{s}\sum_{r}\sum_{\text{ridge}}\big(\tau_{s,r,\text{ridge}}^{\text{true}}-\tau_{s,r,\text{ridge}}^{\text{iter}}\big)^{2}.\label{tt_misfit_def}
\end{equation}
This definition of the travel time misfit includes both small and large-distance measurements with equal weights, in line with the spirit of full-waveform inversions. We do not explore the ramifications of different choices in the misfit function in this work.

An update in the iterated stream function $\psi$ results in a change in the travel-time misfit $\chi$ through
\begin{equation}\label{tt_misfit_psi_relation}
\delta\chi=\int d\mathbf{x}K_{\psi}(\mathbf{x})\delta\psi(\mathbf{x}),
\end{equation}
where ${K_{\psi}(\mathbf{x})}$ is a sensitivity kernel that maps the modelled parameter --- the stream function in this case --- to the observed travel times. It may be viewed as the gradient of the wave travel-time misfit with respect to the stream function. We compute these kernels using the adjoint method \citep{Hanasoge2011} and project them onto the B-spline basis to obtain a relationship between travel-time misfit and model coefficients,
\begin{equation}\label{tt_misfit_decomposed_psi_relation}
\begin{split}\delta\chi & =\sum_{\ell}\sum_{i=0}^{m-1}\bigg[\int d\mathbf{x}K_{\psi}(\mathbf{x})B_{i}(z)f_{\ell}(x)\bigg]\delta c_{i\ell}^{\text{iter}}\\
 & =\sum_{\ell}\sum_{i=0}^{m-1}\delta c_{i\ell}^{\text{iter}}K_{i\ell}.
\end{split}
\end{equation}
$K_{i\ell}$ are the components of the kernel in the spline basis. Gradient in hand, we utilize a suitable optimization scheme such as Broyden-Fletcher-Goldfarb-Shanno algorithm (BFGS) or nonlinear conjugate gradient (CG) \citep{Nocedal2006} to iteratively update the coefficients corresponding to the B-spline functions peaking beneath the surface.

\begin{figure*}[t]
\centering \includegraphics[width=18cm]{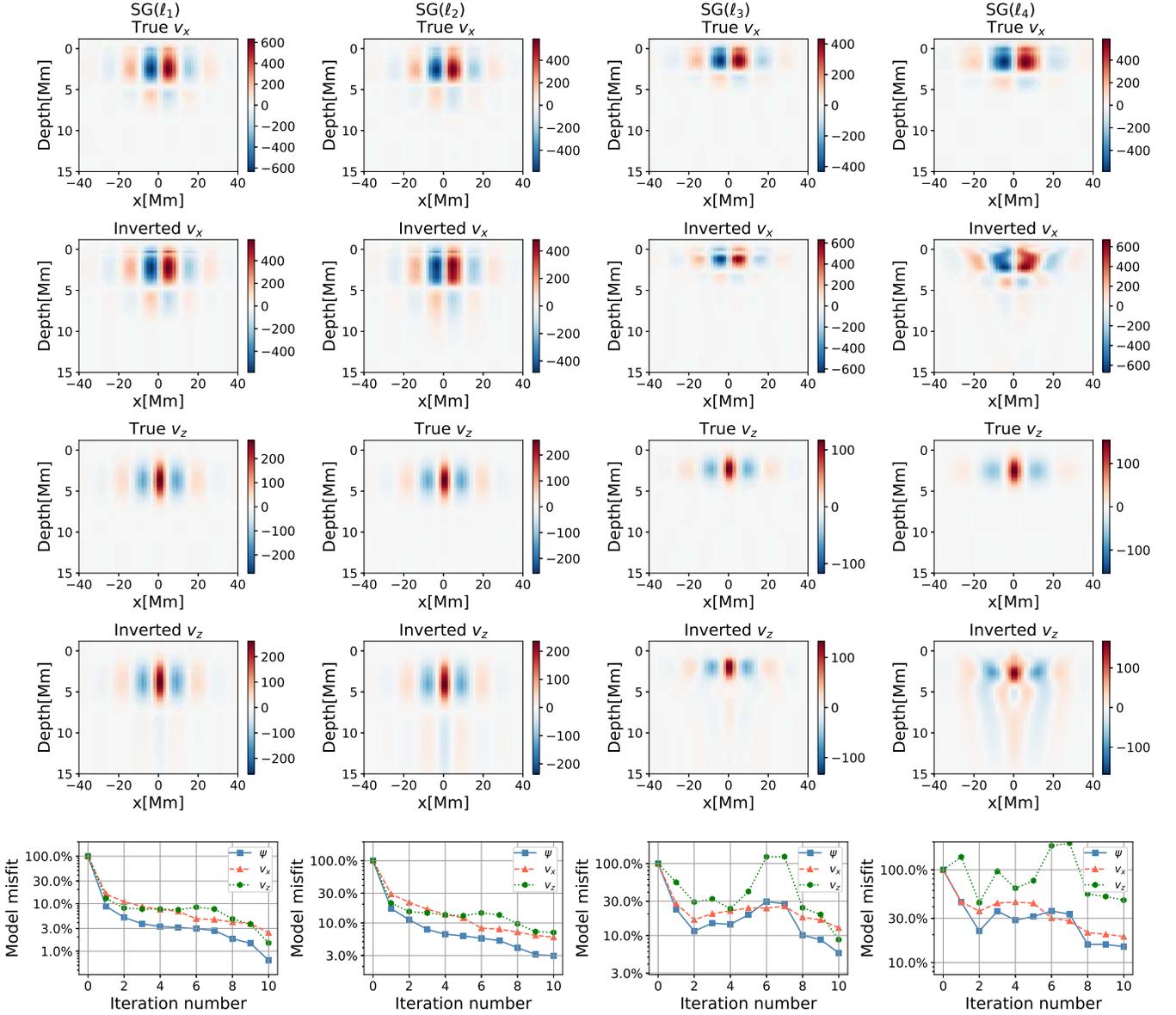}
\caption{True and inverted flow velocities of models SG($\ell_{1}$) - SG($\ell_{4}$)
and model misfits. Each column corresponds to one model. The \textit{topmost panel} in each row indicates the true $v_{x}$; the \textit{second from top panel} indicates the inverted $v_{x}$; the \textit{third panel} indicates the true $v_{z}$; the \textit{fourth panel} indicates the inverted $v_{z}$; and the \textit{bottom-most panel} indicates the model misfits for all flow quantities (Eq. \eqref{model_misfit_def})}
\label{diff_ells_plots} 
\end{figure*}

\section{Results and Discussion}

\begin{figure}[t]
\centering \includegraphics[width=8.5cm]{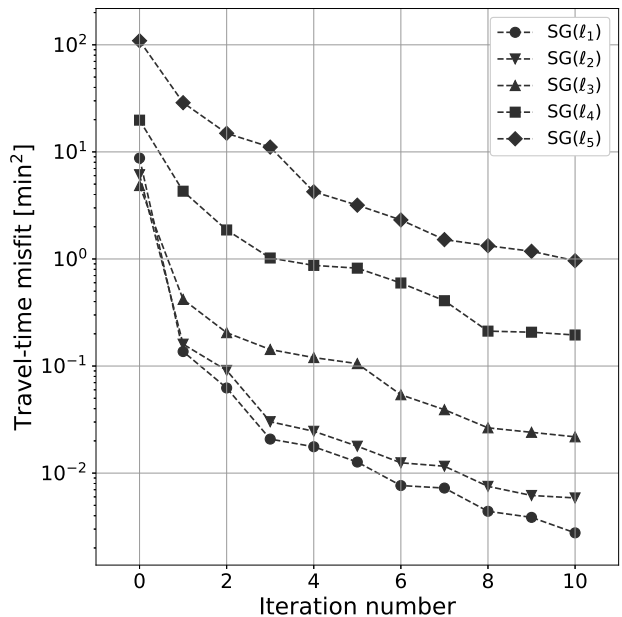}
\caption{Travel-time misfit vs number of iterations for models SG($\ell_{1}$)
- SG($\ell_{5}$) as calculated by Eq. \eqref{tt_misfit_def}}
\label{tt_diff_ells} 
\end{figure}
\subsection{Case 1: Different range of angular degrees}

\label{case1}

\citet{Ferret2019} highlights the incompatibility of averaged $2$D
separable supergranule models with observations and demonstrates the necessity of non-separable models to be able to reproduce velocity observations for an average supergranule. Inversion of a non-separable model poses several challenges, a major one being the large number of parameters. A larger parameter set results in a highly complex parameter space and the likelihood of the optimization scheme to successfully converge to the true model drastically reduces. In Case $1$, we consider flow models with increasing numbers of terms in the stream function, i.e., the range of degrees in Legendre polynomials increases. Models SG($\ell_{1}$) - SG($\ell_{5}$) are supergranule flow models peaking at approximately the same depth (Table \ref{model_parameters}). We quantify the success of the inversion scheme by defining model misfits, $\kappa$, \citep{2017A&A...607A.129B} that indicate the degree to which the iterative model matches the true model
\begin{equation}
\kappa_{\psi}=\frac{\int d\mathbf{x}(\psi^{\text{true}}(\mathbf{x})-\psi^{\text{iter}}(\mathbf{x}))^{2}}{\int d\mathbf{x}(\psi^{\text{true}}(\mathbf{x})-\psi^{\text{start}}(\mathbf{x}))^{2}}.\label{model_misfit_def}
\end{equation}
Similarly, model misfits may be defined for the velocity components.

We implicitly assume that the true and iterative models have the same range of angular degrees and carry out $10$ iterations in each case. We plot the true and inverted velocity profiles and the model misfits of models SG($\ell_{1}$) - SG($\ell_{4}$) in Fig. \ref{diff_ells_plots}, and the travel-time misfit for the five models at the end of each iteration in Fig. \ref{tt_diff_ells}. An inspection of flow profiles and model misfits indicates that while models SG($\ell_{1}$) and SG($\ell_{2}$) --- which contain fewer than 200 parameters --- progressively approach the true model, the models SG($\ell_{3}$) - SG($\ell_{5}$) appear to veer off, despite the travel time misfits from Fig \ref{tt_diff_ells} indicating a similar degree of improvement for all the models. For the last three models, we observe an increase in model misfits which may be arising due to the large number of inversion parameters. The continuous reduction in travel-time misfit for models SG($\ell_{3}$) - SG($\ell_{5}$) hints that we might be converging to a local minimum or saddle point in the parameter space.

\begin{figure*}[t]
\centering \includegraphics[width=16cm]{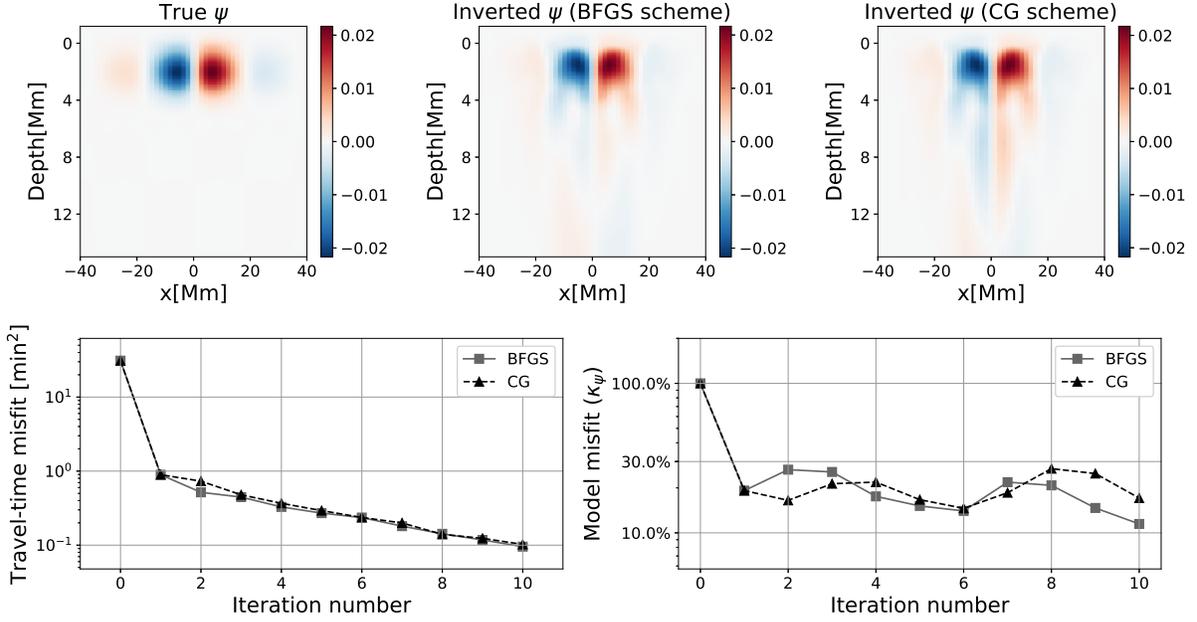}
\caption{True and inverted stream functions for model SG($d_{1}$). \textit{Top-left panel} depicts true $\psi$, \textit{top-center panel} depicts the inverted $\psi$ using the BFGS scheme, and \textit{top-right panel} depicts the inverted $\psi$ using the CG scheme. \textit{Bottom-left panel} plots travel-time misfit for each iteration and \textit{bottom-right panel} plots model misfit at each iteration.}
\label{sgd1_plots} 
\end{figure*}

\subsection{Case 2: Supergranule models peaking at different depths}
Although we are unable to accurately recover the true flow for models with more than a few angular degrees, we ask an alternate question: is it possible to recover the depth of the supergranule? The exact definition of the depth of a supergranule is uncertain, and various authors have used different measures in the past as estimators.  \citet{Duvall1998,Zhao2003} used the depth at which the subsurface flow becomes uncorrelated with the surface velocity, although \citep{Braun2004, Woodard2007} suggested that detecting such a layer might be a challenge, even if it were to exist. In any case this definition might be inaccurate if the horizontal scale of the flow velocity were to vary with depth \citep{Svanda2013ApJ...775....7S}, or potentially misleading if the flow is not temporally stationary \citep{Greer2016ApJ...824..128G}. In our work, we use the depth at which the horizontally averaged squared stream function reaches a maximum as an estimate of the peak depth of a supergranule, defined as
\begin{equation}
\bar{d}=\text{argmax}_z\bigg(\int\psi^{2}(\mathbf{x})dx\bigg).\label{supergranule_depth_def}
\end{equation}
This layer --- if it were to exist --- is closer to the surface and possibly does not suffer from the aforementioned shortcomings. While this is not the actual depth of the supergranule, it may serve as a lower bound.

We construct four flow models, peaking at different depths (Table \ref{model_parameters}): SG($d_{1}$) - SG($d_{4}$), and allow the iterative model to fit for a larger range of angular degrees than the true model. We remove the assumption we had in \ref{case1} and furnish little {\it a priori} information to the inversion algorithm. We perform two sets of inversions for each model, employing the BFGS and CG schemes respectively and find that both methods converge to the same  model, different from the true flow pattern. These results, along with the observation that there is a continual decrease in travel-time misfit suggests that we may have converged to a local minimum. We plot results for the model SG($d_{1}$) in Fig. \ref{sgd1_plots}.

\begin{figure*}[t]
\centering \includegraphics[width=17cm]{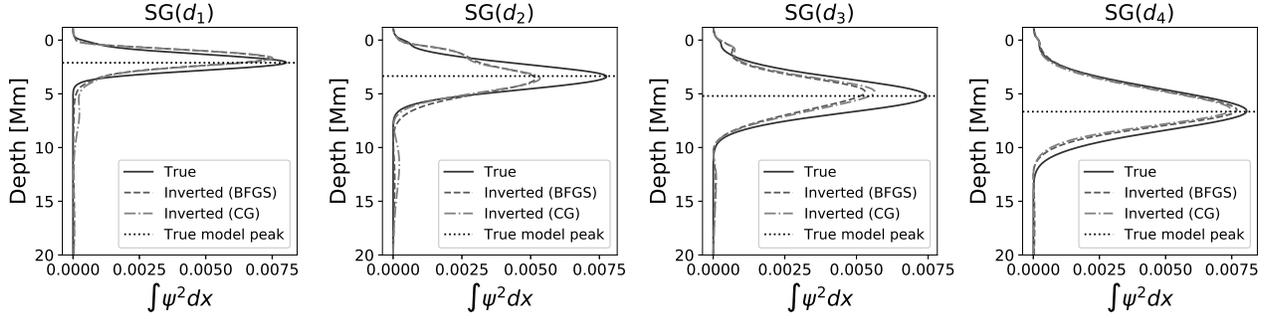} \caption{Depth profiles of models SG($d_{1}$) - SG($d_{4}$). The dotted line corresponds to the peak depth of the particular model.}
\label{depth_profiles} 
\end{figure*}

\begin{figure*}[t]
\centering \includegraphics[width=14cm]{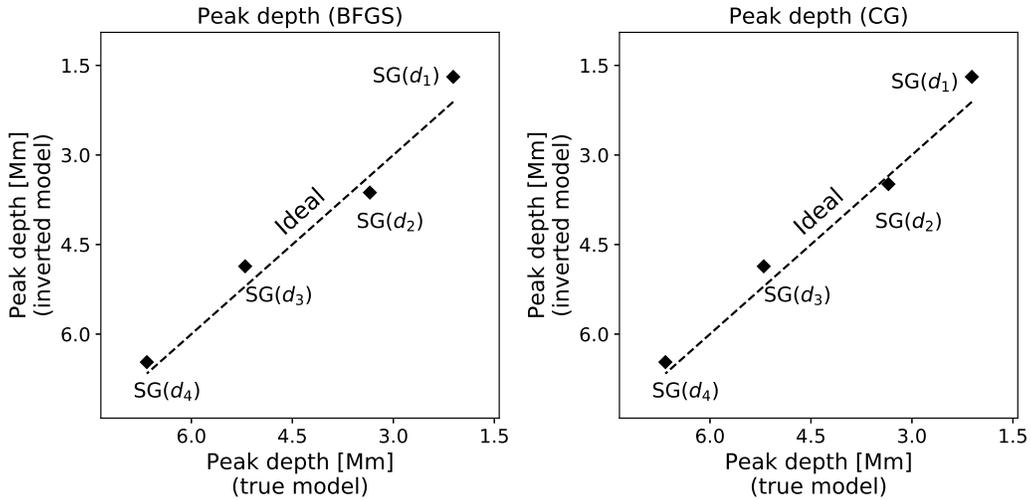}
\caption{Peak depths of models SG($d_{1}$) - SG($d_{4}$). The \textit{left panel} shows the peak depth of the true models and the corresponding inverted model that has been obtained using the BFGS scheme while the \textit{right panel} shows a similar plot where the inversion is carried out using the CG scheme.}
\label{peak_depths} 
\end{figure*}

\begin{figure*}[t]
\centering
\includegraphics[scale=0.47]{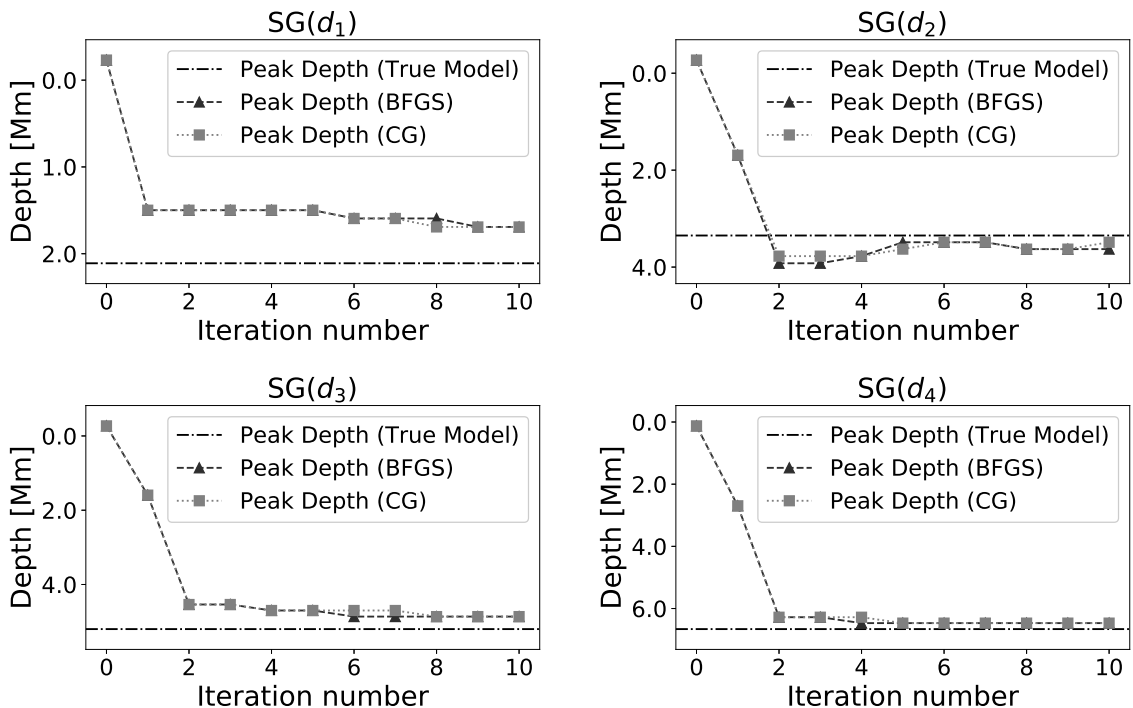}
\caption{Peak depth of iterated models SG($d_{1}$) - SG($d_{4}$) at the end of each iteration.}
\label{peak_depth_v_iteration}
\end{figure*}

We show the variation of the horizontally averaged squared stream function along the vertical axis for models SG($d_{1}$) - SG($d_{4}$) in Fig. \ref{depth_profiles} and observe that the profile for the inverted models peaks close to that of the true model. This is highlighted in Fig. \ref{peak_depths}, where we plot the peak depths \eqref{supergranule_depth_def} and achieve a semblance accuracy. We plot the progression of peak depth of the iterative updated flow models in Fig. \ref{peak_depth_v_iteration} and we observe that it converges to that of the true model in nearly two iterations for all cases. It is encouraging that we are able to replicate comparable values for the peak depth in spite of not recovering the full flow models accurately.

\section{Conclusion}

The success of a high-dimensional optimization scheme often depends on the exact type of regularization imposed on the solution. Choices such as Tikhonov regularization \citep{Dombroski2013} or Fourier smoothing \citep{Bhattacharya2016} have been implemented in the past, but such approaches have often failed to converge to the global minimum - corresponding to the true solution. In this work, we have chosen to follow \citet{2017A&A...607A.129B} and impose an implicit regularization by expressing our velocity fields in a smooth basis of horizontal Legendre polynomials and vertical B-splines. We demonstrate that with this choice of regularization travel-time inversions for non-separable models of supergranules are able to recover their peak depths accurately. Further work is necessary to establish the extent to which this result holds in the more realistic scenario of noisy measurements. It is expected that the signal-to-noise will improve by a factor of $\sqrt{N}$ on averaging over $N$ supergranule cells, but this might still limit the depth to which the sensitivity kernels can probe \citep{Dombroski2013}. Our current result indicates that it might be possible to set bounds based on the signal-to-noise level. It would be interesting to probe how such a limit derived from time-distance seismology compares with that derived from mode-coupling \citep{Woodard2007} or holographic estimates \citep{Braun2007}. Additionally we might need to include a model-covariance matrix \textit{a-priori} \citep[i.e., in parameter space][]{Tarantola1982RvGSP..20..219T}. In this work we sidestep this by choosing a basis that automatically introduces such a correlation through its functional form, although it might be possible to obtain better estimates using simulations of solar convection. More importantly, including the data and model covariance matrices would allow us to compute uncertainties on the inferred profile, something that is lacking in the current analysis. Such an approach will facilitate checking for consistency in the various inferred results.

\bibliographystyle{aasjournal}

\bibliography{references}

\end{document}